\documentclass[twocolumn,letter]{jpsj3}

\usepackage{color}
\usepackage{bm}

\title{%
Regularized Framework of a Weyl Equation for Describing a Weyl Semimetal:
Application to the Case with a Screw Dislocation
}

\author{%
Yositake Takane
}

\inst{%
Department of Quantum Matter, Graduate School of Advanced Sciences of Matter,\\
Hiroshima University, Higashihiroshima, Hiroshima 739-8530, Japan
}

\recdate{ \hspace{50mm} }

\abst{%
The term Weyl semimetal originates from the fact
that its energy dispersion obeys a Weyl equation.
However, a Weyl equation itself cannot fully describe
the electron states in an actual bounded geometry.
For example, the appearance of chiral surface states, which is a characteristic
feature of a Weyl semimetal, cannot be captured with a Weyl equation.
This indicates that some degree of freedom is lost
when a Weyl equation is derived from a microscopic model of a Weyl semimetal.
To overcome this difficulty, we present a framework
consisting of a Weyl equation and a supplementary equation,
which can be derived from a microscopic model.
Applying this framework to a cylindrical system
in the presence of a screw dislocation,
we show that it appropriately describes the chiral surface states
and one-dimensional chiral modes along a dislocation line.
The local charge current induced by these chiral states
is determined in an analytical manner.
}

\begin{document}
\sloppy
\maketitle

The valence and conduction bands of a Weyl semimetal
touch each other at pairs of isolated points, around which
the band structure obeys a Weyl equation.~\cite{murakami,
wan,yang,burkov1,burkov2,WK,delplace,halasz,sekine,weng,huang1,xu1,lv1,lv2,xu2}
That is, the band structure around each band-touching point (i.e., a Weyl node)
is equivalent to a nondegenerate three-dimensional (3D) Dirac cone.
Bulk electron states consisting of 3D Dirac cones are called Weyl states.
A pair of Weyl nodes has opposite chirality.
We hereafter restrict our consideration to the case
with only a single pair of Weyl nodes at $\mib{k}_{+}$ and $\mib{k}_{-}$.
A notable feature of a Weyl semimetal is that two-dimensional (2D) gapless
states with chirality appear on its surface.~\cite{wan}
The presence of such chiral surface states gives rise to
an anomalous Hall effect.~\cite{burkov1}
Even in the interior of a Weyl semimetal, gapless states can appear
in the presence of a screw dislocation (see Fig.~1)
if it is parallel to the line connecting the Weyl nodes.~\cite{imura1}
The gapless states localized along a dislocation line
propagate in only one direction; thus, they are referred to as
one-dimensional (1D) chiral modes.
This is in contrast to the helical nature of 1D modes realized
in a weak topological insulator.~\cite{ran,imura2}

The term Weyl semimetal originates from the fact that
its energy dispersion obeys a Weyl equation.
However, a Weyl equation itself cannot capture the 1D and 2D chiral states.
Furthermore, even bulk Weyl states cannot be described by a Weyl equation
in an actual bounded geometry.
Since a Weyl equation can be derived from a microscopic model for
a Weyl semimetal, this suggests that significant information
about the electron states is partially lost in its derivation.
The purpose of this letter is to present a simple framework that enables us to
describe the electron states in a bounded Weyl semimetal
on the basis of a Weyl equation.
Decomposing the eigenvalue equation for a microscopic model
of a Weyl semimetal, we derive a Weyl equation and a supplementary equation.
We then show that if the Weyl equation is combined with
the supplementary equation, it appropriately describes
the 1D and 2D chiral states and bulk Weyl states.
\begin{figure}[btp]
\begin{center}
\includegraphics[height=2.0cm]{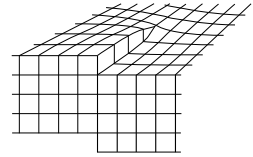}
\end{center}
\caption{
Screw dislocation with a displacement of
one unit atomic layer (i.e., $N = 1$).
}
\end{figure}

Another purpose is to analyze the response of a Weyl semimetal
upon the insertion of a screw dislocation.
Recently, Sumiyoshi and Fujimoto~\cite{sumiyoshi} showed that
a local charge current appears near a dislocation center
owing to the fictitious magnetic field~\cite{kawamura}
induced by the dislocation.
The crucial point is that the local current near the dislocation center
is finite, although total charge current vanishes.
This phenomenon is interesting, as it can be regarded as
a chiral magnetic effect in the static limit.~\cite{zyuzin,chen,vazifeh,
chang,yamamoto,takane1,ibe}
However, as the analysis in Ref.~\citen{sumiyoshi} basically relies on
a Weyl equation, the 1D chiral modes along a screw dislocation
are not taken into consideration.
Such 1D chiral modes should more significantly contribute
to the charge current than the bulk Weyl states.
We examine the phenomenon by applying a framework consisting of
the Weyl and supplementary equations
and find an analytical expression for the local charge current
arising from the 1D chiral modes and 2D chiral surface states.
We set $\hbar = 1$ hereafter.

To start, we derive the Weyl and supplementary equations
from a microscopic model for a Weyl semimetal.
We assume that the wavenumber $k_{z}$ in the $z$ direction
is a good quantum number
and take the continuum limit in the $x$ and $y$ coordinates,
whereas the $z$ coordinate is discretized with the lattice constant $a$.
Then, the microscopic model of Ref.~\citen{yang} is
\begin{align}
   H = \left[ 
         \begin{array}{cc}
           \Delta(k_{z})+B(\hat{k}_{x}^{2}+\hat{k}_{y}^{2})
              & A(\hat{k}_{x}-i\hat{k}_{y}) \\
           A(\hat{k}_{x}+i\hat{k}_{y})
              & -\Delta(k_{z})-B(\hat{k}_{x}^{2}+\hat{k}_{y}^{2})
         \end{array}
       \right] ,
\end{align}
where $\hat{k}_{x}=-i\partial_{x}$, $\hat{k}_{y}=-i\partial_{y}$, and
\begin{align}
  & \Delta(k_{z}) = -2t\bigl[\cos(k_{z}a)-\cos(k_{0}a)\bigr] .
\end{align}
Here, $0< k_{0} < \frac{\pi}{a}$, and $A$, $B$, and $t$
are assumed to be real and positive.
The Weyl nodes are located at $\mib{k}_{\pm} = (0,0,\pm k_{0})$,
where the energy is zero.
We assume the absence of an external magnetic field,
although our argument allows a uniform magnetic field.
Note that if $B=0$, the eigenvalue equation of $H\Psi = E\Psi$
is reduced to a Weyl equation.~\cite{comment1}
We show that the eigenvalue equation is decomposed into two equations
unless $B = 0$.
Expressing $\Psi(x,y)$ as $\Psi = {}^t\!(F,G)$, we easily find
\begin{align}
   \left[ B^{2}\mathcal{D}_{\parallel}^{2}
          -(A^{2}+2B\Delta)\mathcal{D}_{\parallel}
          +\Delta^{2}-E^{2} \right]F = 0
\end{align}
with $\mathcal{D}_{\parallel} = \partial_{x}^{2}+\partial_{y}^{2}$.
A solution of this equation must satisfy either
$\mathcal{D}_{\parallel}F=\Lambda_{-}F$
or $\mathcal{D}_{\parallel}F=\Lambda_{+}F$, where
\begin{align}
  \Lambda_{\pm} = \frac{A^{2}+2B\Delta\pm\sqrt{(A^{2}+2B\Delta)^{2}
                        +4B^{2}(E^{2}-\Delta^{2})}}{2B^{2}} .
\end{align}
The same statement also holds for $G$.
This indicates that the eigenvalue equation has two different families
of solutions; $F$ and $G$ respectively satisfy
$\mathcal{D}_{\parallel}F = \Lambda_{-}F$
and $\mathcal{D}_{\parallel}G = \Lambda_{-}G$ in the first family and
$\mathcal{D}_{\parallel}F = \Lambda_{+}F$
and $\mathcal{D}_{\parallel}G = \Lambda_{+}G$ in the second family.
For both families, the eigenvalue equation requires
\begin{align}
  \left(\Delta-E-B\mathcal{D}_{\parallel}\right)F
  -iA (\partial_{x}-i\partial_{y})G = 0 .
\end{align}
Hereafter, we restrict our consideration to the case
where $B$ is very small but finite, in which $\Lambda_{\pm}$ is reduced to
\begin{align}
          \label{eq:def-Lambda}
  \Lambda_{-} = -\frac{E^{2}-\Delta^{2}}{A^{2}} ,
         \hspace{6mm}
  \Lambda_{+} = \frac{A^{2}}{B^{2}} .
\end{align}
The equation with $\Lambda_{-}$ is equivalent to the Weyl equation.
The equation with $\Lambda_{+}$ is referred to as a supplementary equation,
which disappears if $B=0$ from the outset.
The latter yields only solutions that exponentially increase or decrease.
Our argument indicates that an eigenfunction is constructed by
superposing the solutions of both the Weyl and supplementary equations.

To show how this framework captures a 2D chiral surface
state,~\cite{okugawa,ramamurthy,takane2}
we apply it to the system occupying the region of $x \ge 0$
with the surface at $x = 0$.
The wavenumbers $k_{y}$ and $k_{z}$ are good quantum numbers.
The Weyl equation allows an exponentially decreasing function of $x$
only when $|\Delta(k_{z})| > |E|$,
which is necessary to construct a localized surface state.
Then, the general solution is
\begin{align}
        \label{eq:ge-solu}
  \Psi(x)
   = c
     \left[ \begin{array}{c}
              \gamma-k_{y} \\
              i\frac{\Delta-E}{A}
            \end{array}
     \right]
     e^{-\gamma x}
   + d
     \left[ \begin{array}{c}
              1 \\
              -i
            \end{array}
     \right]
     e^{-\kappa x} ,
\end{align}
where $c$ and $d$ are arbitrary constants, and
\begin{align}
  \gamma = \sqrt{\frac{\Delta^{2}-E^{2}}{A^{2}}+k_{y}^{2}} ,
         \hspace{7mm}
  \kappa = \frac{A}{B} .
\end{align}
The first and second terms of Eq.~(\ref{eq:ge-solu}) respectively arise
from the Weyl and supplementary equations.
The boundary condition of $\Psi(0)={}^t\!(0,0)$ requires
$\gamma-k_{y} = -(\Delta-E)/A$.
This holds only when $\Delta(k_{z}) < 0$,
indicating that 2D chiral surface states appear in the region of
$\Delta(k_{z}) < E < -\Delta(k_{z})$ with $k_{z}\in (-k_{0}, k_{0})$.
Solving the equation,
we find that $E = -Ak_{y}$ independent of $k_{z}$.
This state propagates only in the $-y$ direction,
which is a characteristic feature of the 2D chiral surface states.

Let us start the analysis of the electron states
in the presence of a screw dislocation.
We consider a cylindrical Weyl semimetal of radius $R$ and length $L_{z}$
under the periodic boundary condition in the $z$ direction,
by using the cylindrical coordinates $(r,\phi,z)$
with $r = \sqrt{x^{2}+y^{2}}$ and $\phi = \arctan(y/x)$.
A screw dislocation parallel to the $z$ axis
is inserted at $r = 0$ with a displacement of $N$ unit atomic layers.
That is, its Burgers vector is $\mib{b} = a(0,0,N)$.
It is convenient to rewrite $F$ and $G$ of $\Psi(r,\phi) = {}^t\!(F,G)$
as $F = e^{il\phi} f(r)$ and $G = e^{i(l+1)\phi} g(r)$,
where $l$ is the azimuthal quantum number.
The screw dislocation is described by
the effective vector potential:~\cite{kawamura,comment2}
$e(A_{r},A_{\phi},A_{z}) = (0,\zeta(k_{z})/r,0)$ with
\begin{align}
   \zeta(k_{z}) = \frac{Na}{2\pi}k_{z} .
\end{align}
The Weyl equation for $f$ and $g$ is given by
\begin{align}
        \label{eq:f-first}
 \left(\mathcal{D}_{l}-\Lambda_{-}\right)f = 0 ,
        \hspace{3mm}
 \left(\mathcal{D}_{l+1}-\Lambda_{-}\right)g = 0 ,
\end{align}
while the supplementary equation is
\begin{align}
        \label{eq:f-second}
 \left(\mathcal{D}_{l}-\Lambda_{+}\right)f = 0 ,
        \hspace{3mm}
 \left(\mathcal{D}_{l+1}-\Lambda_{+}\right)g = 0 ,
\end{align}
where $\Lambda_{-}$ and $\Lambda_{+}$ are given in Eq.~(\ref{eq:def-Lambda}),
and
\begin{align}
      \label{eq:def-D}
 \mathcal{D}_{l} = \partial_{r}^{2}+\frac{1}{r}\partial_{r}
                     - \frac{(l+\zeta)^{2}}{r^{2}} .
\end{align}
The eigenvalue equation requires that
\begin{align}
      \label{eq:suppl-Dirac}
  \left(\Delta-E-B\mathcal{D}_{l}\right)f
  -iA\bigl(\partial_{r}+\frac{l+1+\zeta}{r}\bigr)g = 0 .
\end{align}
As $l$ is a good quantum number, we express an eigenfunction
for a given $l$ as $\Psi(r) = {}^t\!(f, g)$.
In the absence of a screw dislocation, our system consists of
2D circular electron systems of radius $R$ stacked in the $z$ direction;
thus, the appropriate boundary condition is $\Psi(R) = {}^t\!(0,0)$.
However, once a screw dislocation is inserted at $r = 0$,
it becomes topologically equivalent to a set of $N$ helix staircases.
Hence, we should impose $\Psi(0) = {}^t\!(0,0)$
in addition to $\Psi(R) = {}^t\!(0,0)$.

We present solutions of the two equations
to construct the general solution of the eigenvalue equation
assuming that $\zeta$ is not an integer.
The Weyl equation is equivalent to a modified Bessel equation
when $|\Delta|>|E|$
and is equivalent to a Bessel equation when $|E|>|\Delta|$.
With $\eta \equiv \sqrt{\Delta^{2}-E^{2}}/A$,
the solutions when $|\Delta|>|E|$ are given by
\begin{align}
    \psi_{l+\zeta}^{\eta}(r) & =
    {}^t\!\bigl[I_{l+\zeta}(\eta r), iR'(E)I_{l+1+\zeta}(\eta r)\bigr] ,
       \\
    \psi_{-l-\zeta}^{\eta}(r) & =
    {}^t\!\bigl[I_{-l-\zeta}(\eta r), iR'(E)I_{-l-1-\zeta}(\eta r)\bigr] ,
\end{align}
where $R'(E)=(-\Delta+E)/\sqrt{\Delta^{2}-E^{2}}$.
Here, $I_{\nu}(x)$ is the $\nu$th-order modified Bessel function
of the first kind.
With $q \equiv \sqrt{E^{2}-\Delta^{2}}/A$,
the solutions when $|E|>|\Delta|$ are given by
\begin{align}
   \psi_{l+\zeta}^{q}(r) & =
   {}^t\!\bigl[J_{l+\zeta}(q r), iR(E)J_{l+1+\zeta}(q r)\bigr] ,
          \\
   \psi_{-l-\zeta}^{q}(r) & =
   {}^t\!\bigl[J_{-l-\zeta}(q r), -iR(E)J_{-l-1-\zeta}(q r)\bigr] ,
\end{align}
where $R(E)=(E-\Delta)/\sqrt{E^{2}-\Delta^{2}}$.
Here, $J_{\nu}(x)$ is the $\nu$th-order Bessel function of the first kind.
Let us turn to the supplementary equation, which is equivalent to
a modified Bessel equation independent of $E$.
With $\kappa \equiv A/B$, the solutions are given by
\begin{align}
   \psi_{l+\zeta}^{\kappa}(r) & =
   {}^t\!\bigl[I_{l+\zeta}(\kappa r), iI_{l+1+\zeta}(\kappa r)\bigr] ,
        \\
   \psi_{-l-\zeta}^{\kappa}(r) & =
   {}^t\!\bigl[I_{-l-\zeta}(\kappa r), iI_{-l-1-\zeta}(\kappa r)\bigr] .
\end{align}

Now, we determine the energy dispersion of the electron states.
Let us first consider the case of $|\Delta(k_{z})|>|E|$, in which
the solutions of the Weyl equation are written
by using modified Bessel functions.
These functions and their linear combinations asymptotically increase or
decrease in an exponential manner, and should describe spatially localized
states (i.e., 1D chiral modes and 2D chiral surface states).
The general solution is written as
\begin{align}
  \Psi = c_{1} \psi_{l+\zeta}^{\eta} + d_{1} \psi_{l+\zeta}^{\kappa}
         + c_{2} \psi_{-l-\zeta}^{\eta}
         + d_{2} \psi_{-l-\zeta}^{\kappa} .
\end{align}
For a given $l$ and $\zeta(k_{z})$, we need to treat the cases
of $l+\zeta > 0$, $l+1+\zeta <0$, and $l+\zeta < 0 < l+1+\zeta$ separately.
In the first case, the boundary condition of $\Psi(0)={}^t\!(0,0)$ is satisfied
by setting $c_{2} = d_{2} = 0$.
Imposing the other boundary condition of $\Psi(R)={}^t\!(0,0)$, we find
\begin{align}
          \label{eq:disp-CS1}
  \frac{-\Delta+E}{\sqrt{\Delta^{2}-E^{2}}}
     = \frac{I_{l+\zeta}(\eta R)}{I_{l+1+\zeta}(\eta R)} ,
\end{align}
where $I_{\nu}(\kappa R)/I_{\nu+1}(\kappa R) \approx 1$ is used.
Equation~(\ref{eq:disp-CS1}) holds only when $\Delta(k_{z}) < 0$,
as the right-hand side is positive.
When $l+\zeta \ll \eta R$, we approximately obtain
$E = (A/R)(l+\zeta+1/2)$,~\cite{imura1} which is rewritten as
\begin{align}
       \label{eq:disp-CS_state}
  E = \frac{ANa}{2\pi R}\left(k_{z}-\tilde{k}_{l}\right)
\end{align}
with
\begin{align}
      \label{eq;def-tilde_k_l}
  \tilde{k}_{l} = \frac{2\pi}{Na}\left(-l -\frac{1}{2} \right).
\end{align}
As both $\psi_{l+\zeta}^{\eta}(r)$ and $\psi_{l+\zeta}^{\kappa}(r)$
exponentially increase with increasing $r$,
the resulting $\Psi(r)$ must be localized near the surface at $r = R$.
Therefore, Eq.~(\ref{eq:disp-CS_state}) should be identified as
the energy dispersion of a 2D chiral surface state.
Our argument indicates that such a state appears only in the region of
$\Delta(k_{z}) < E < -\Delta(k_{z})$.
Similarly, the energy in the case of $l+1+\zeta <0$ is determined by
\begin{align}
           \label{eq:disp-CS2}
  \frac{-\Delta+E}{\sqrt{\Delta^{2}-E^{2}}}
     = \frac{I_{-l-\zeta}(\eta R)}{I_{-l-1-\zeta}(\eta R)} ,
\end{align}
which again leads to Eq.~(\ref{eq:disp-CS_state}) when $-l-\zeta \ll \eta R$.
In the last case of $l+\zeta < 0 < l+1+\zeta$, both a 1D chiral mode and
2D chiral surface state appear, as seen in the following.
Imposing $\Psi(R)={}^t\!(0,0)$ and $\Psi(0)={}^t\!(0,0)$, we find
\begin{align}
         \label{eq:disp-CS3}
  & R'(E)I_{l+1+\zeta}(\eta R)-I_{l+\zeta}(\eta R)
       \nonumber \\
  & + \left(\frac{\eta}{\kappa}\right)^{1+2l+2\zeta}
      \left[\frac{I_{-l-\zeta}(\eta R)}{R'(E)}-I_{-l-1-\zeta}(\eta R)\right]
    = 0 .
\end{align}
Let us focus on the case with $\eta R \gg 1$,
where Eq.~(\ref{eq:disp-CS3}) is decomposed into
\begin{align}
  R'(E) = \frac{I_{l+\zeta}(\eta R)}{I_{l+1+\zeta}(\eta R)} ,
                   \hspace{5mm}
  R'(E) = \left(\frac{\eta}{\kappa}\right)^{1+2l+2\zeta} .
\end{align}
The first equation is essentially equivalent to Eq.~(\ref{eq:disp-CS1}),
which determines the energy dispersion of the 2D chiral surface state.
The second equation holds only when $\Delta(k_{z}) < 0$,
under which it is reduced to
\begin{align}
        \label{eq:1D-chiral}
  E = |\Delta| -\frac{2|\Delta|}
                     {1+e^{-2(1+2l+2\zeta)\ln\left(\kappa/\eta\right)}} .
\end{align}
Note that $E$ becomes zero at $\zeta = -\frac{1}{2} - l$
(i.e., $k_{z} = \tilde{k}_{l}$).
Near this point, Eq.~(\ref{eq:1D-chiral}) results in
\begin{align}
       \label{eq:1D-chiral-dis}
  E = -\frac{|\Delta(\tilde{k}_{l})|Na}{\pi}
       \ln\left(\frac{A^{2}}{B|\Delta(\tilde{k}_{l})|}\right)
       (k_{z}-\tilde{k}_{l}) ,
\end{align}
which should be identified as the energy dispersion of a 1D chiral mode.
Here, let us count the number of 1D chiral modes, $n_{c}$,
for a given $k_{0}$ and $N$.
The condition of $\Delta(k_{z}^{l}) < 0$ indicates that
$n_{c}$ is equal to the total number of values of $l$
that satisfy $-k_{0} < \tilde{k}_{l} < k_{0}$.
For example, $n_{c}$ in the case of $N = 2$ is $2$ if $k_{0} > \frac{\pi}{2a}$
and $0$ if $k_{0} < \frac{\pi}{2a}$.

We next consider the case of $|E| > |\Delta(k_{z})|$,
in which only bulk Weyl states appear.
The general solution is
\begin{align}
  \Psi = c_{1} \psi_{l+\zeta}^{q} + d_{1} \psi_{l+\zeta}^{\kappa}
         + c_{2} \psi_{-l-\zeta}^{q}
         + d_{2} \psi_{-l-\zeta}^{\kappa} ,
\end{align}
to which $\Psi(0)={}^t\!(0,0)$ and $\Psi(R)={}^t\!(0,0)$ are imposed.
In the case of $l+\zeta > 0$, the energy is determined by
\begin{align}
          \label{eq:disp-bW1}
  \frac{E-\Delta}{\sqrt{E^{2}-\Delta^{2}}}
     = \frac{J_{l+\zeta}(q R)}{J_{l+1+\zeta}(q R)} .
\end{align}
The energy in the case of $l+1+\zeta <0$ is determined by
\begin{align}
          \label{eq:disp-bW2}
  \frac{E-\Delta}{\sqrt{E^{2}-\Delta^{2}}}
     = - \frac{J_{-l-\zeta}(q R)}{J_{-l-1-\zeta}(q R)} .
\end{align}
In the last case of $l+\zeta < 0 < l+1+\zeta$, we find that
\begin{align}
         \label{eq:disp-bW3}
  & R(E)J_{l+1+\zeta}(q R)-J_{l+\zeta}(q R)
       \nonumber \\
  & - \left(\frac{q}{\kappa}\right)^{1+2l+2\zeta}
      \left[\frac{J_{-l-\zeta}(q R)}{R(E)}+J_{-l-1-\zeta}(q R)\right]
    = 0 ,
\end{align}
which determines the energy of the bulk Weyl states.

The energy dispersion is fully determined by
Eqs.~(\ref{eq:disp-CS1}), (\ref{eq:disp-CS2}), (\ref{eq:disp-CS3}),
(\ref{eq:disp-bW1}), (\ref{eq:disp-bW2}), and (\ref{eq:disp-bW3}).
We now comment on the crossover of a 1D (or 2D) chiral state
to a bulk Weyl state.
On the plane spanned by $E$ and $k_{z}$, chiral states appear
in the region of $\Delta(k_{z}) < E < -\Delta(k_{z})$
with $k_{z} \in (-k_{0},k_{0})$,
whereas bulk Weyl states appear outside the region.
Note that $\Delta(k_{z}) < 0$ for $k_{z} \in (-k_{0},k_{0})$.
The two regions are separated by the two lines of
$E = \mp \Delta(k_{z})$ [see the dotted lines in Fig.~2(a)].
Below, they are referred to as the upper and lower boundary lines.
A 1D (or 2D) chiral state tends to extend spatially
when approaching either of the boundary lines
and is continuously connected to a bulk Weyl state at some point on the line.

\begin{figure}[btp]
\begin{center}
\includegraphics[width=5.4cm]{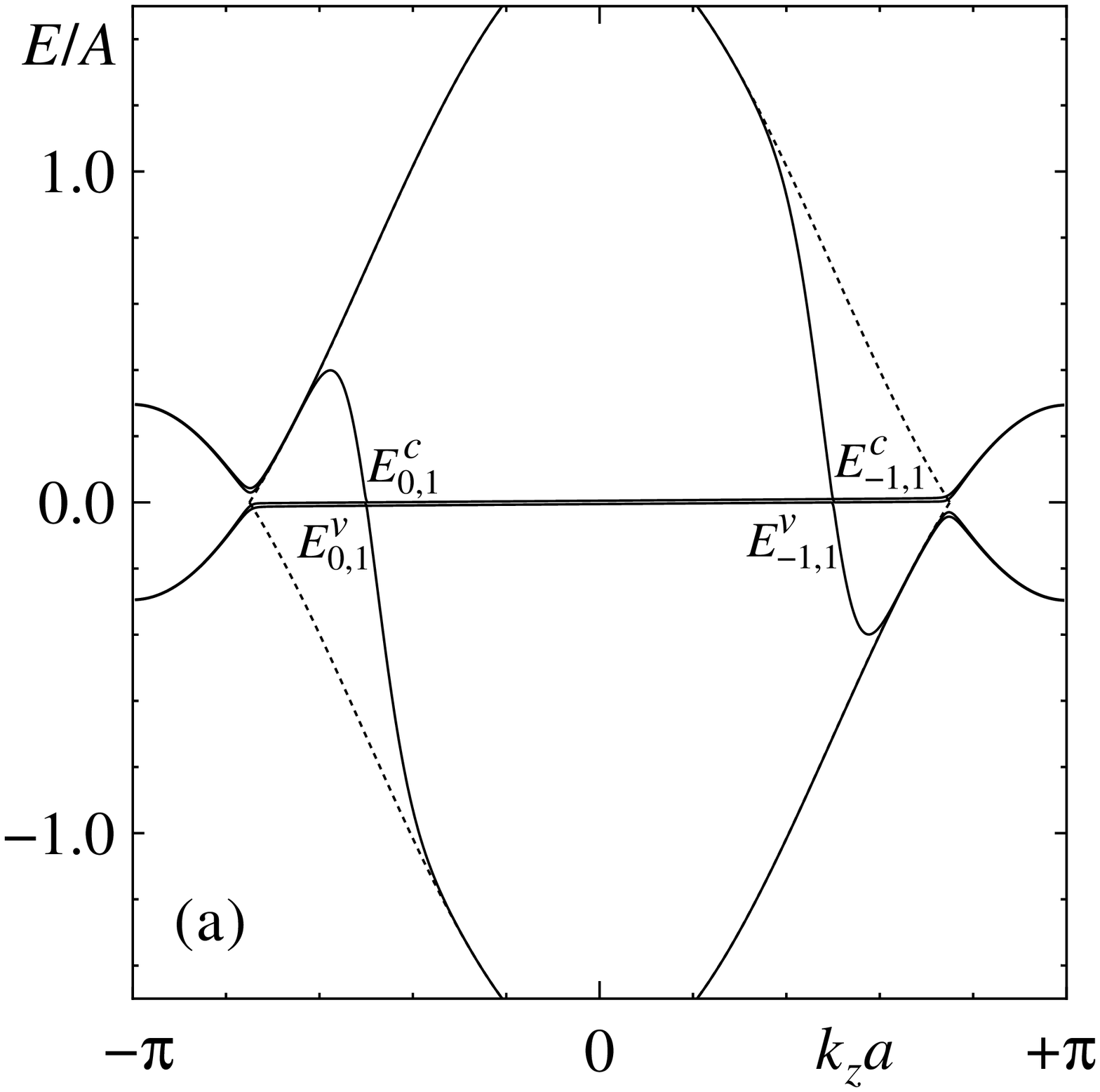}
\includegraphics[width=5.5cm]{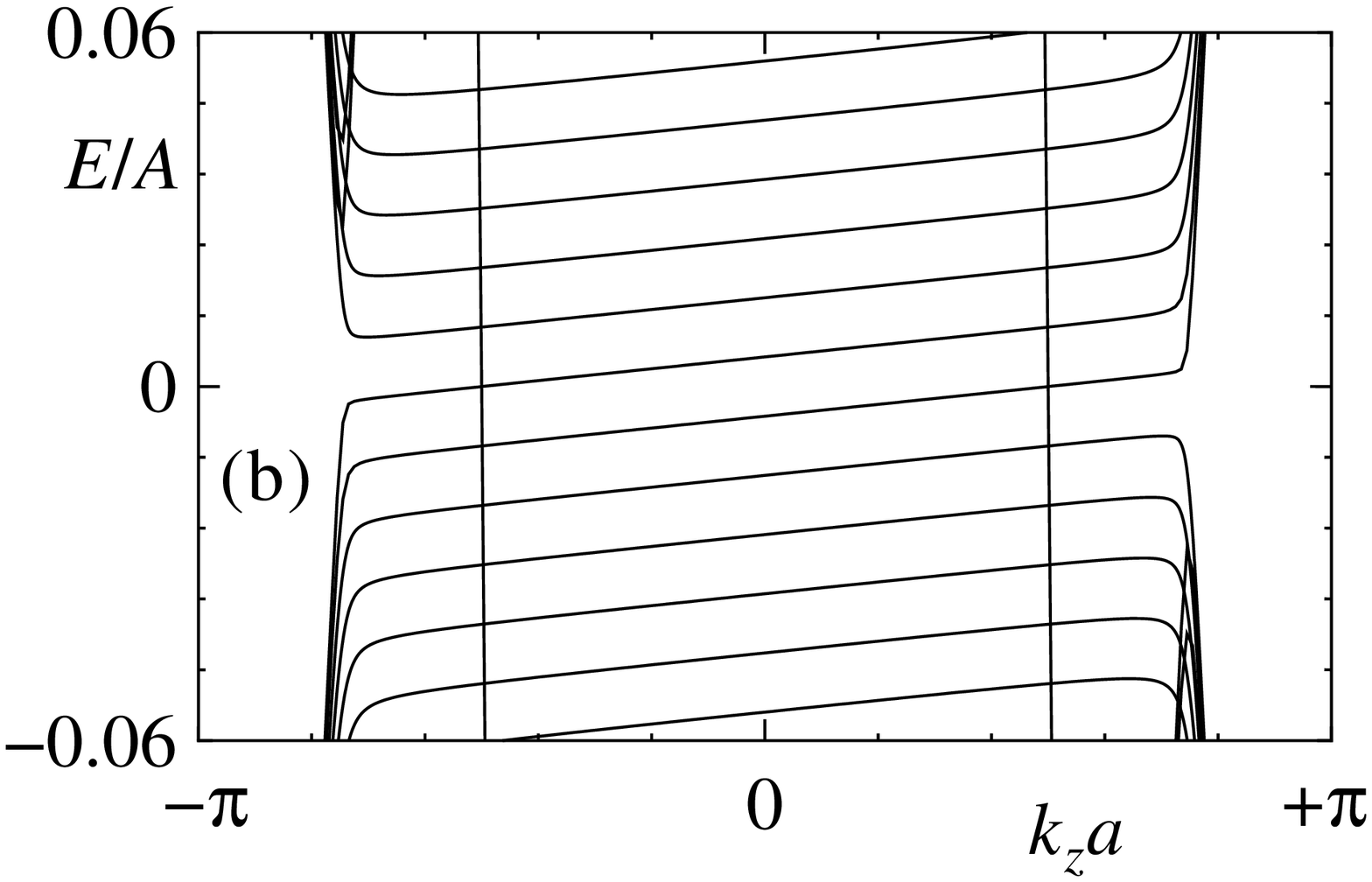}
\end{center}
\caption{
Energy dispersion in the case of $N = 2$.
(a) $E_{0,1}^{v}$, $E_{0,1}^{c}$, $E_{-1,1}^{v}$, and $E_{-1,1}^{c}$
(solid lines) and the boundary lines (dotted lines).
(b) $E_{l,1}^{v}$ for $l = 0$ to $-6$ and $E_{l,1}^{c}$ for $l = -1$ to $5$.
}
\end{figure}
Let $E_{l,m}^{v}(k_{z})$ and $E_{l,m}^{c}(k_{z})$ be the energies
of the $m$th state with $l$ in the valence and conduction bands,
respectively, as a function of $k_{z}$, where $m = 1$, $2$, $\dots$ is assigned
in descending order (i.e., $0 > E_{l,1}^{v} > E_{l,2}^{v} > \dots$)
in the former and in ascending order
(i.e., $0 < E_{l,1}^{c} < E_{l,2}^{c} < \dots$) in the latter.
As an example, we determine the energy dispersion of several states
in the case of $N = 2$ and $k_{0} = \frac{3\pi}{4a}$, for which $n_{c} = 2$,
with $t/(Aa^{-1}) = 0.5$, $B/(Aa) = 0.001$, and $R/a = 100$.
Figure~2 shows the results for $m = 1$.
The states with $m \ge 2$ are bulk Weyl states.
In Fig.~2(a), the solid lines represent $E_{0,1}^{v}$, $E_{0,1}^{c}$,
$E_{-1,1}^{v}$, and $E_{-1,1}^{c}$, while dotted lines represent
the lower and upper boundary lines.
The 1D chiral modes with $l = 0$ and $-1$ respectively appear near
$k_{z} = -\frac{\pi}{2a}$ and $\frac{\pi}{2a}$.
The 2D chiral surface states with $l = 0$ and $-1$ appear
near $E = 0$ with a nearly flat dispersion.
They almost overlap.
Figure~2(b) shows $E_{l,1}^{v}$ for $l = 0$ to $-6$
and $E_{l,1}^{c}$ for $l = -1$ to $5$, representing
a series of the 2D chiral surface states with $l = -6$ to $5$.
Nearly vertical lines represent the 1D chiral modes with $l = 0$ and $-1$.

Let us consider the local charge current along the $z$ direction.
This is dominated by $J_{\rm 1D}$ and $J_{\rm 2D}$, which respectively
represent the contribution from the 1D chiral modes localized near $r = 0$ and
that from the 2D chiral surface states localized near $r = R$.
Bulk Weyl states, which extend over the entire system,
also contribute to the local charge current.
However, from an experimental viewpoint,
their contribution is less important than $J_{\rm 1D}$ and $J_{\rm 2D}$,
as it is thinly distributed over the entire cross section of the system,
reflecting the extended nature of the bulk Weyl states.
In the analysis given below, the Fermi level is located
at the Weyl nodes (i.e., $E_{\rm F} = 0$).
We focus on the 1D chiral mode with $l$,
the energy of which decreases with increasing $k_{z}$
and crosses the line of $E = 0$ at $k_{z} = \tilde{k}_{l}$.
This mode is eventually connected to a bulk Weyl state at some point,
$\tilde{k}_{l}^{\rm max}$, on the lower boundary line.
The contribution to $J_{\rm 1D}$ arises from occupied states within
$\tilde{k}_{l} < k_{z} \lesssim \tilde{k}_{l}^{\rm max}$.
Noting that the energy dispersion of the 1D mode is given
by $E_{l,1}^{v}(k_{z})$ as long as $E < 0$, we find that
\begin{align}
  \frac{J_{\rm 1D}}{L_{z}}
  = -\frac{e}{2\pi} \sum_{l}{}^{'} 
     \int_{\tilde{k}_{l}}^{\tilde{k}_{l}^{\rm max}} dk_{z}
     \frac{\partial E_{l,1}^{v}}{\partial k_{z}} ,
\end{align}
where the summation over $l$ is restricted by
$-k_{0} < \tilde{k}_{l} < k_{0}$.
Approximately replacing $\tilde{k}_{l}^{\rm max}$ with $\tilde{k}_{l}$,
we find that~\cite{comment3}
\begin{align}
         \label{eq:J-1D}
  \frac{J_{\rm 1D}}{L_{z}}
  = \frac{e}{2\pi}
    \sum_{l}{}^{'} |\Delta(\tilde{k}_{l})| .
\end{align}
This indicates that $J_{\rm 1D}$ is mainly determined by $\Delta(k_{z})$
and the $N$ dependence manifests itself
through the restricted summation over $l$.
Let us turn to the derivation of $J_{\rm 2D}$.
The group velocity of the 2D chiral surface states is $v=ANa/(2\pi R)$
and independent of $l$ and $k_{z}$; thus,
$J_{\rm 2D}$ is determined by the number of occupied states.
Note that the 2D chiral surface states are almost uniformly distributed over
the region encircled by the boundary lines (see Fig.~2).
From this observation, we find that
\begin{align}
         \label{eq:J-2D}
  \frac{J_{\rm 2D}}{L_{z}}
  = -e\frac{Na}{4\pi^{2}}\Sigma(k_{0}) ,
\end{align}
where
\begin{align}
  \Sigma(k_{0}) = \int_{-k_{0}}^{k_{0}}dk_{z}|\Delta(k_{z})| .
\end{align}
Note that the direction of flow is opposite to that of $J_{\rm 1D}$.

Let us finally consider the total charge current along the $z$ direction.
It must vanish,~\cite{vazifeh,sumiyoshi} although the local current notably
appears in the two regions near $r = 0$ and $r = R$.
That is, the charge current induced by the bulk Weyl states completely cancels
$J_{\rm 1D}$ and $J_{\rm 2D}$
if they are integrated over the entire cross section.
This is simply observed by using an extended zone representation
of electron states.
To avoid complexity, we focus on the case with $N = 2$.
As $E_{\rm F} = 0$, we consider only the valence band in the following.
The Weyl and supplementary equations ensure that $E_{l,m}^{v}(k_{z})$
defined for $k_{z} \in [-\frac{\pi}{a}, \frac{\pi}{a}]$
satisfies $E_{l,m}^{v}\left(\frac{\pi}{a}\right)
= E_{l+2,m}^{v}\left(-\frac{\pi}{a}\right)$ for any $l$ and $m$.
Thus, we classify $\{E_{l,m}^{v}(k_{z}):l \in \mathbb{Z}\}$ into
two subsets:
$\{E_{l,m}^{v}(k_{z}):l \in 2 \mathbb{Z}\}$
and $\{E_{l,m}^{v}(k_{z}):l \in 2 \mathbb{Z}+1\}$.
Each subset is fully represented by a single function
defined for $k_{z} \in (-\infty, \infty)$.
Indeed, we can define
$\epsilon_{m}^{v,i}(k_{z})$ with $k_{z}\in (-\infty, \infty)$
for $i = e$, $o$ by
$\epsilon_{m}^{v,e}(k_{z}+\frac{\pi l}{a}) = E_{l,m}^{v}(k_{z})$
for $l \in 2 \mathbb{Z}$ and
$\epsilon_{m}^{v,o}(k_{z}+\frac{\pi(l+1)}{a}) = E_{l,m}^{v}(k_{z})$
for $l \in 2 \mathbb{Z}+1$, where $k_{z} \in [-\frac{\pi}{a}, \frac{\pi}{a}]$.
Note that $\epsilon_{m}^{v,e}(k_{z})$ and $\epsilon_{m}^{v,o}(k_{z})$
are continuous functions of $k_{z} \in (-\infty, \infty)$
and describe all of the electron states in the valence band,
including the 1D and 2D chiral states and bulk Weyl states.
As $E_{l,m}^{v}(k_{z}) \to -\infty$ in the limit of $l \to \pm\infty$
for a given $k_{z}$, $\epsilon_{m}^{v,i}(k_{z}) \to -\infty$
if $k_{z} \to \pm\infty$.
Since the spectrum is unbounded,
we introduce a cutoff at $E = -E_{c}$ with $E_{c} > 0$.
Then, the total charge current is expressed as
\begin{align}
  \frac{J_{\rm total}}{L_{z}}
  = - \frac{e}{2\pi} \sum_{i=e,o}\sum_{m=1}^{\infty}
      \int_{-\infty}^{\infty}dk_{z}
      \Theta\left(\epsilon_{m}^{v,i}+E_{c}\right)
      \frac{\partial \epsilon_{m}^{v,i}}{\partial k_{z}} ,
\end{align}
where $\Theta(x)$ is the Heaviside function.
Applying the argument given in the supplemental material
of Ref.~\citen{sumiyoshi}, we can show that $J_{\rm total} = 0$
irrespective of $E_{c}$.

\section*{Acknowledgment}

This work was supported by JSPS KAKENHI Grant Number 15K05130.


\begin{thebibliography}{99}
\bibitem{comment1} The eigenvalue equation at $B = 0$ reads
\begin{align}
   \left[ 
      \begin{array}{cc}
        \Delta(k_{z}) & A(\hat{k}_{x}-i\hat{k}_{y}) \\
        A(\hat{k}_{x}+i\hat{k}_{y}) & -\Delta(k_{z})
      \end{array}
  \right] \Psi
  = E \Psi ,
       \nonumber
\end{align}
which is reduced to a Weyl equation
if $\Delta(k_{z})$ is linearized at $k_{z} = \pm k_{0}$.
\bibitem{comment2} The vector potential creates a fictitious magnetic field
at $r = 0$; thus, one may think that it breaks the assumption of
our framework that a magnetic field is absent.
However, this is not the case thanks to the boundary condition of
$\Psi(0) = {}^t\!(0,0)$.
\bibitem{comment3} It is meaningful to rewrite Eq.~(\ref{eq:J-1D}) as
\begin{align}
 \frac{J_{\rm 1D}}{L_{z}}
  = e\frac{Na}{4\pi^{2}}\times
    \frac{2\pi}{Na}\sum_{l}{}^{'}|\Delta(\tilde{k}_{l})| ,
          \nonumber
\end{align}
where the second factor in the right-hand side can be regarded as
a rough approximation of $\Sigma(k_{0})$.
This indicates that $J_{\rm 1D}$ is on the same order of $J_{\rm 2D}$.
\end{thebibliography}
\end{document}